\pdfoutput=1
\documentclass[twocolumn,prb,letterpaper,showpacs,superscriptaddress]{revtex4-2}
\usepackage{amsmath}
\usepackage{graphicx}
\graphicspath{{Figures/}}
\usepackage{amssymb}
\usepackage{color}
\usepackage{siunitx}
\DeclareSIUnit \belm {Bm}
\DeclareSIUnit \sample {Sa}
\newcommand{\pq}[2]{#1\,{\rm #2}}% physical quantity, value and unit
\newcommand{\Tn}{T_{\rm N}}% Noise temperature
\newcommand\nuJ{\nu_\mathrm{J}}
\newcommand\nuA{\nu_\mathrm{in}}
\newcommand\nuB{\nu_\mathrm{out}}

\newcommand\Ohm{\Omega}
\usepackage[modulo]{lineno}
\begin{document}

\title{Microwave photon-number amplification}

\author{R.~Albert}
\affiliation{Univ.~Grenoble Alpes, CEA, INAC-PHELIQS, F-38000 Grenoble, France}

\author{J.~Griesmar}
\affiliation{Institut Quantique, Université de Sherbrooke, Sherbrooke, Québec J1K 2R1, Canada}

\author{F.~Blanchet}
\affiliation{Univ.~Grenoble Alpes, CEA, INAC-PHELIQS, F-38000 Grenoble, France}

\author{U.~Martel}
\affiliation{Institut Quantique, Université de Sherbrooke, Sherbrooke, Québec J1K 2R1, Canada}

\author{N.~Bourlet}
\affiliation{Institut Quantique, Université de Sherbrooke, Sherbrooke, Québec J1K 2R1, Canada}

\author{M.~Hofheinz}
\affiliation{Institut Quantique, Université de Sherbrooke, Sherbrooke, Québec J1K 2R1, Canada}
\affiliation{Univ.~Grenoble Alpes, CEA, INAC-PHELIQS, F-38000 Grenoble, France}

\date{\today}
\begin{abstract}
  So far, quantum-limited power meters are not available in the microwave domain, hindering measurement of photon number in itinerant quantum states. On the one hand, single photon detectors \cite{chen11, kono18, besse18, royer18, lescanne_detecting_2020, grimsmo21} accurately detect single photons, but saturate as soon as two photons arrive simultaneously. On the other hand, more linear watt meters, such as bolometers \cite{richards94, karasik11, lee20}, are too noisy to accurately detect single microwave photons. Linear amplifiers \cite{yurke_observation_1989-1, bergeal_phase-preserving_2010-1, macklin_nearquantum-limited_2015} probe non-commuting observables of a signal so that they must add noise \cite{caves_quantum_1982} and cannot be used to detect single photons, either.   Here we experimentally demonstrate a microwave photon-multiplication scheme which combines the advantages of a single photon detector and a power meter by multiplying the incoming photon number by an integer factor. Our first experimental implementation achieves a $n=3$-fold multiplication with 0.69 efficiency in a $\SI{116}{\mega\hertz}$ bandwidth up to a input photon rate of $\SI{400}{MHz}$. It loses phase information but does not require any dead time or time binning. We expect an optimised device cascading such multipliers to achieve number-resolving measurement of itinerant photons with low dark count, which would offer new possibilities in a wide range of quantum sensing and quantum computing applications.
\end{abstract}

\maketitle

\begin{figure}[t]
  \includegraphics[width=\columnwidth]{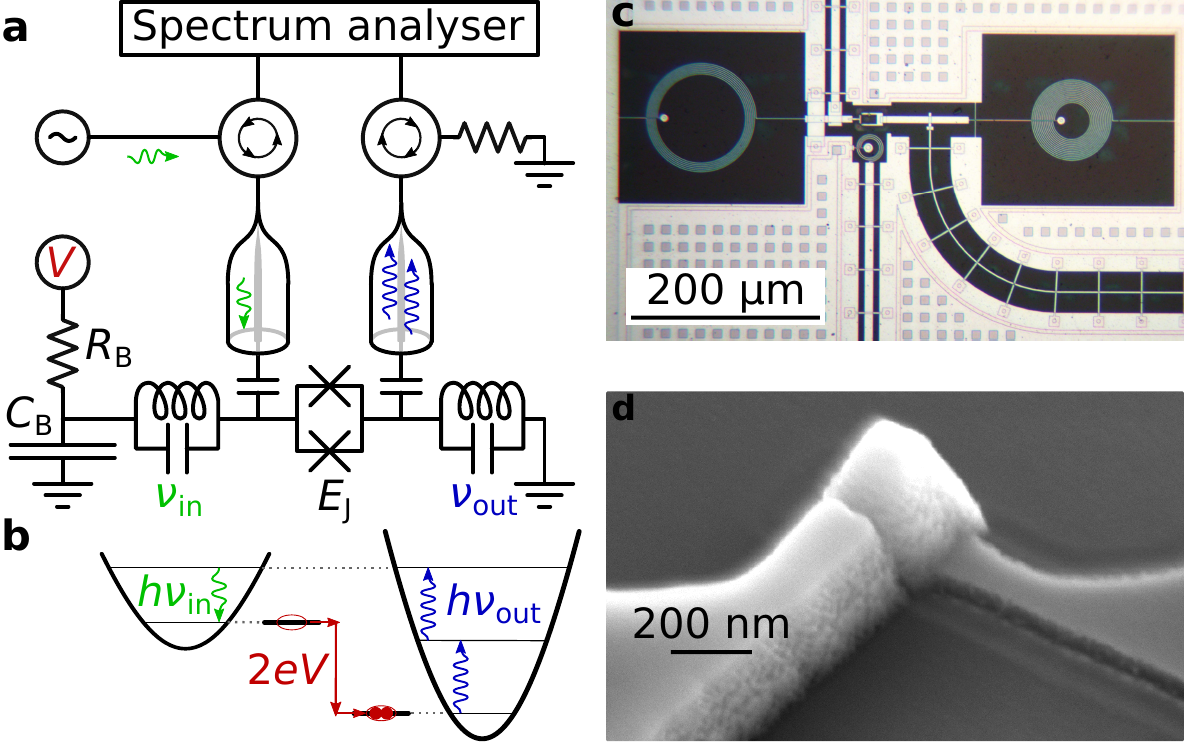}
  \caption{
    Setup, sample and working principle.
    \textbf{a}, The sample consists of two buffer resonators at frequencies $\nuA = \SI{4.8}{\giga\hertz}$ and $\nuB = \SI{6.13}{\giga\hertz}$. The resonators are non-linearly coupled by a SQUID biased at a voltage $V$ via a heavily filtered bias line. Its source impedance is modelled by a $R_B = \pq{5}{\Ohm}$ resistor at an effective temperature of $\pq{90}{mK}$ (see App. \ref{app:vnoise}). An on-chip capacitor $C_B \approx \pq{100}{pF}$ shunts $R_B$  at the operation frequency (see App. \ref{app:fullsetup}).
    \textbf{b}, The voltage bias $V$ is set such that $h\nuA + 2eV = n h \nuB$. Under this condition an incoming photon in mode $a$ is converted into $n$ photons in mode $b$, with the energy $2eV$ of a tunnelling Cooper pair providing the required energy.
    \textbf{c}, Optical micrograph of the photon-multiplier. The large coils define the inductance of the input and output modes, the small one at the centre provides a flux bias to tune the SQUID.
    \textbf{d}, Scanning electron micrograph of one of the $\SI{170}{nm}$ by $\SI{170}{nm}$ Nb-Al-AlOx-Nb junctions of the SQUID.
    \label{fig:setup}}
\end{figure}

Quantum measurements of propagating microwaves are so far mostly performed with linear amplifiers. If such an amplifier is able to amplify signals irrespectively of their phase, it must add noise. This noise is due to the fact that the quadratures of electromagnetic fields do not commute and, therefore, cannot be accurately measured at the same time. In order to respect the commutation relations of incoming and outgoing fields, the amplifier must couple the signal mode to an additional idler mode \cite{haus62, caves_quantum_1982}. The zero-point fluctuations of this mode then appear as additional noise in the amplified signal mode. The added noise can be avoided if the amplification is made phase-sensitive, i.e.\ only one quadrature is amplified and the other attenuated by the same factor \cite{caves_quantum_1982}, so that phase-space volume is preserved. Therefore, linear amplifiers add noise or are blind to one quadrature of the signal and, therefore, cannot be used to discriminate single photons from vacuum.

This discrimination can be achieved with single photon detectors which have recently become available in the microwave domain \cite{chen11, kono18, besse18, royer18, lescanne_detecting_2020, grimsmo21}. They allow measuring the power of very weak signals without added photon noise by discarding phase information rather than one of the quadratures. But single-photon detectors are intrinsically strongly non-linear, with their binary outcome  mapping all incoming non-zero photon number states to the same output state. This non-invertible behaviour discards more information than strictly required by commutation relations, in particular the actual photon number. Many implementations also require frequent resets or use of a latching mechanism, so that even more information on the incoming photon state is lost, e.g.\ its actual arrival time.

In this work we experimentally demonstrate a linear photon-number amplification scheme. It deamplifies phase-information, so that, like for a phase-sensitive amplifier, phase-space volume is preserved and no noise needs to be added. It allows to optimally measure the intensity of signals with unknown phase, such as single photons, making it complementary to phase sensitive amplifiers which are optimal for measuring signals with well-defined phase reference. A single photon detector represents a limiting case to our scheme with infinite multiplication factor, so that any incoming photon saturates the amplifier. Similar number-resolving detectors have so far only be implemented for photons residing in cavities \cite{peaudecerf14, essig21}, which allow for repeated measurements of the state.\\

We implement this photon-number amplifier using Josephson photonics \cite{hofheinz11, souquet16, leppakangas_multiplying_2018, leppakangas18, rolland19, peugeot21, menard22} based on inelastic Cooper pair tunnelling \cite{ingold92, holst94}. Our device consists of a voltage-biased Josephson junction coupled to input and output lines via two low-Q cavities with resonance frequencies $\nuA$ and $\nuB$, see Fig.~\ref{fig:setup}a. The Josephson junction is biased at a voltage $V$ such that $2eV + h\nuA = n h\nuB$. Under this condition a photon in the input mode at $\nuA$ can be converted into $n$ photons in the output mode at $\nuB$, with a tunnelling Cooper pair providing the difference in energy $2eV$, as seen in Fig.~\ref{fig:setup}b \cite{leppakangas_multiplying_2018}. In contrast to usual linear amplifiers where the gain depends on a continuously variable tuning parameter, the gain $n$ is an integer number set by the bias condition. The integer gain allows the scheme to be noiseless and phase preserving by having a phase-space response with $n$-fold rotational symmetry, so that phase-space volume is preserved.  The photo-multiplication effect we describe corresponds, for $n=1$, to the parametric frequency-conversion mode with unity gain of a parametric amplifier, which appears when the pump is set at the difference between idler and signal frequency. 

The experimental setup is presented in Fig.~\ref{fig:setup}a. The two low-Q cavities are realised by spiral $LC$ resonators visible in the optical micrograph of Fig.~\ref{fig:setup}c. Their resonances are centred at $\nuA = \SI{4.8}{\giga\hertz}$ and $\nuB = \SI{6.13}{\giga\hertz}$ and have characteristic impedances $Z_c \approx \SI{400}{\Ohm}$. A SQUID acts as a Josephson junction with Josephson energy $E_J(\Phi)$, tunable in situ by a local magnetic flux $\Phi$. While we can not directly measure its critical current, measurements of test junctions on the same chip indicate a typical critical current $I_c \approx \SI{60}{nA}$. Fig.~\ref{fig:setup}d shows a SEM picture of one of the Nb-Al-AlOx-Nb junctions of the SQUID. The flux line is low-pass filtered at dilution temperature with a homemade Eccosorb filter~\cite{paquette22}. A DC voltage bias is applied to the SQUID via a $\SI{5}{\Ohm}/\SI{1}{\mega\Ohm}$ voltage divider with heavy low-pass filtering \cite{albert_multiplication_2019}. The two $LC$ resonators (input and output) are capacitively coupled to microwave transmission lines. The capacitors are designed to obtain associated coupling rates around $\gamma_\mathrm{in}\approx \gamma_\mathrm{out}\approx\SI{90}{\mega\hertz}$.\\

\begin{figure}
  \includegraphics[width=\columnwidth]{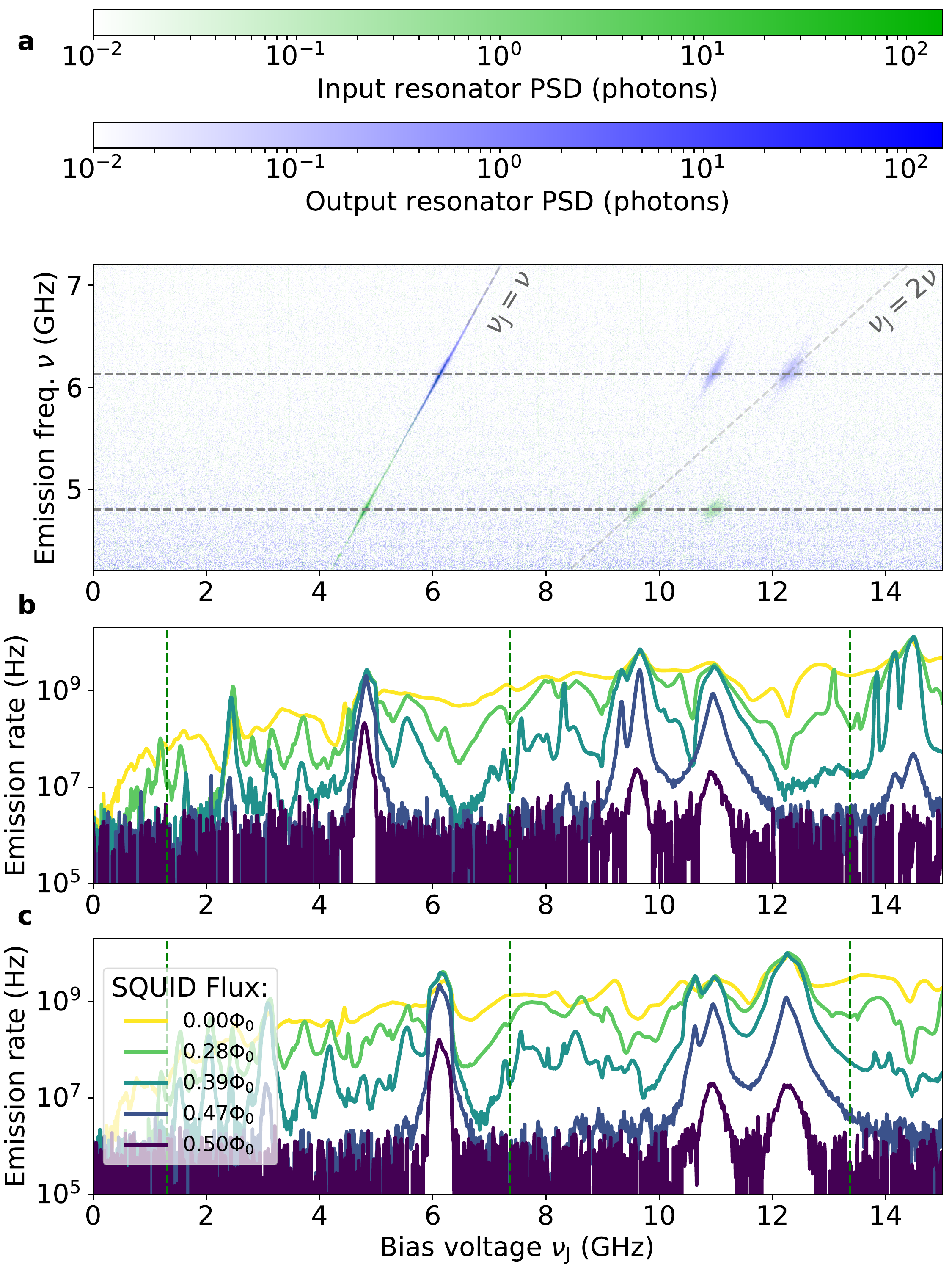}
  \caption{Spontaneous emission.
  \textbf{a}, Power Spectral Density (PSD) as a function of emission frequency and bias voltage. The PSD for the input (output) resonator is plotted in shades of green (blue). The flux in the SQUID loop is set to $\Phi_0/2$ to minimise the Josephson energy of the device.
  \textbf{b} and \textbf{c}, Photon emission rate from the input and output resonator, respectively, integrated over a $\SI{400}{\mega\hertz}$ bandwidth centred at $\SI{4.8}{\giga\hertz}$ and $\SI{6.13}{\giga\hertz}$, respectively, for 5 values of flux in the SQUID loop. These frequencies are represented by horizontal dashed grey lines in \textbf{a}. The flux values of $\SI{0.47}{\Phi_0}$, $\SI{0.39}{\Phi_0}$ and $\SI{0.28}{\Phi_0}$ correspond to the dashed vertical lines in Fig.~\ref{fig:impedancematching} and the flux used in Fig.~\ref{fig:satpower}. The three vertical green dashed lines correspond, from left to right, to the voltages used for the conversions of Fig.~\ref{fig:impedancematching} and~\ref{fig:satpower} with, respectively, $n=1$, $2$ and $3$.
    \label{fig:spontaneous}}
\end{figure}

Fig.~\ref{fig:spontaneous} shows the Power Spectral Density (PSD) of spontaneous emission from the device cooled down to $T\approx\SI{10}{\milli\kelvin}$. For this measurement, no microwave tone is applied to the photo-multiplier. Only the DC voltage bias, expressed in terms of its Josephson frequency $\nuJ=2eV/h$, and the magnetic flux are tuned. At maximal SQUID frustration $\Phi=\Phi_0/2$ (see Fig.~\ref{fig:spontaneous}a), only emission close to the input ($\SI{4.8}{\giga\hertz}$) and output ($\SI{6.13}{\giga\hertz}$) modes is visible. In this panel, the PSD of both the input (blue) and output (green) resonator is represented. The brightest spots around $\nuJ=\SI{4.8}{\giga\hertz}$ and $\SI{6.13}{\giga\hertz}$ are on the $\nuJ=\nu$ line and correspond to the emission of one photon per tunnelling Cooper pair in the corresponding resonator. The emission of two photons per tunnelling Cooper pair is also visible near the $\nuJ=2\nu$ line. The last two spots around $\nuJ=\nuA+\nuB\approx\SI{11}{\giga\hertz}$ correspond to the emission of one photon in each resonator per tunnelling Cooper pair.

Fig.~\ref{fig:spontaneous}b and c show the emission rate of the input and output resonators, integrated over a $\SI{400}{\mega\hertz}$ bandwidth centred at, respectively, the input and output resonance frequencies, indicated by horizontal dashed lines in Fig.~\ref{fig:spontaneous}a. The lines for $\Phi = \Phi_0/2$ show the well-defined processes discussed above. At larger Josephson energies (fluxes closer to $0$), the emission peaks are less well-resolved as more and more complex processes emerge. At the highest Josephson energy, higher order processes involving several Cooper pairs and photons in many modes give rise to emission at nearly all bias voltages.\\

\begin{figure}
  \includegraphics[width=\columnwidth]{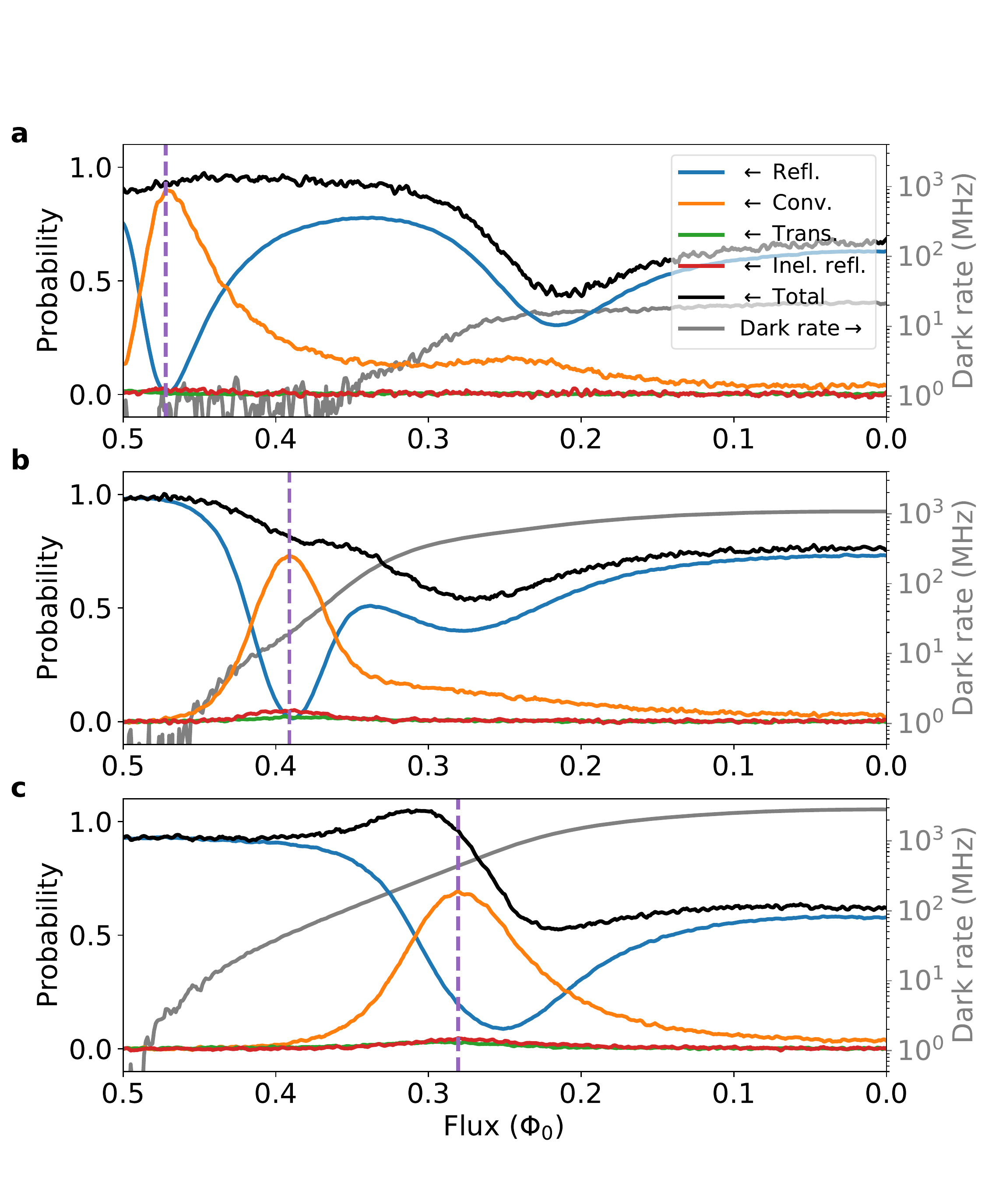}
  \caption{Reflection, conversion, transmission and inelastic reflection probabilities as well as dark count rates for $n=1$ (\textbf{a}), 2 (\textbf{b}) and 3 (\textbf{c}) as a function of the flux in the SQUID loop. The vertical dashed lines represent the flux values at which the conversion probabilities are highest. In \textbf{a}, the input frequency is $\SI{4.773}{\giga\hertz}$ and the bias voltage is $\SI{1.30}{\giga\hertz}$. In \textbf{b}, the input frequency is $\SI{4.71}{\giga\hertz}$ and the bias voltage is $\SI{7.37}{\giga\hertz}$. In \textbf{c}, the input frequency is $\SI{4.74}{\giga\hertz}$ and the bias voltage is $\SI{13.37}{\giga\hertz}$. For these three plots, the input power of $\SI{-127}{\deci\belm}$ is chosen to have approximately one photon on average in the input resonator.
    \label{fig:impedancematching}}
\end{figure}

We then bias the SQUID at voltages $V$ such that $h\nuA + 2eV = n h\nuB$, which enables the multiplication of photons in our device. The voltages corresponding to $n=1$, $2$ and $3$ are indicated by vertical dashed lines in Fig.~\ref{fig:spontaneous}. When a microwave signal at frequency $\nu$ is applied on the input of the device, the desired converted signal appears centred at a frequency $\nu_\mathrm{conv} = \nuB + \left(\nu - \nuA\right)/n$. For $n=1$, its bandwidth is the same as the input signal (modulo the noise coming from DC voltage fluctuations). For $n>1$, the converted photons will be spread over the output resonator bandwidth~\cite{leppakangas_multiplying_2018}. We measure this converted signal by integrating the PSD of the output mode over a $\Delta \nu=\SI{400}{\mega\hertz}$ bandwidth centred around $\nu_\mathrm{conv}$. We also measure the inelastic reflection, i.e.\ leakage of the converted signal through the input port, at the same frequency and bandwidth. The input signal can also be reflected (or transmitted) elastically by the sample. We measure these two elastic signals by integrating the PSD of the input (output) mode over a $\delta \nu=\SI{15}{\mega\hertz}$ bandwidth around $\nu$ to account for phase noise added by the device. Note that, before integration, the PSD of the spontaneous emission is first subtracted from the desired signals and that the PSDs are divided by $h\nu$ to obtain PSDs in term of photon rates densities, as in Fig.~\ref{fig:spontaneous}.

To obtain the probabilities corresponding to those different outcomes, the photon rates obtained after the integration of the different PSDs are divided by the photon rate we apply at the input of the device. In the case of the converted signals, the result is divided by $n$ to account for multiplication. The input rate is calibrated by applying a microwave signal at frequency $\nu$ and integrating the PSD reflected by the sample biased at $\Phi=\Phi_0/2$ and $\nuJ=\SI{2}{\giga\hertz}$, where it is essentially an open circuit reflecting all the input power (see App.~\ref{app:calibration} for more details on the calibration).

These probabilities are presented in Fig.~\ref{fig:impedancematching}, where the input frequency $\nu$ is chosen to obtain a maximum of conversion. The input power is kept low to have, on average, less than one photon in the input mode. The sum of the four probabilities is close to 1 at low Josephson energy (flux close to $\Phi_0/2$), indicating that elastic scattering and the desired conversion process fully explain the device. At larger Josephson energies it drops significantly below 1. The missing signal is likely converted to unmonitored frequencies: with increasing Josephson energy new processes become important \cite{meister15}, involving several Cooper pairs and photons in different spurious modes of the circuit. Especially low-frequency modes, even if they have low quality factor and low characteristic impedance, may then be strongly driven by the Josephson junction and take away arbitrary energy from the conversion process. As seen in Fig.~\ref{fig:spontaneous}, those processes are quite difficult to identify from the measurement background when the Josephson energy and the bias voltage increase. 

The sum of probabilities can, as in Fig.~\ref{fig:impedancematching}c, also exceed 1. Under the bias conditions used here ($\nuJ=\SI{13.37}{\giga\hertz}$, rightmost vertical dashed line in Fig.~\ref{fig:spontaneous}b and c), the spontaneous emission of the device (dark rates) is higher than for lower $n$. This indicates that an incoming microwave signal can also be amplified by stimulated emission \cite{jebari_near-quantum-limited_2018} before being converted or reflected, pushing the observed total probability above 1.

The three conversion probabilities (Fig.~\ref{fig:impedancematching}, orange curves) reach a well-defined maximum as a function of the flux, indicated by vertical dashed lines. At this point, the conversion rate matches the loss rate of the input mode, leading to destructive interference in elastic reflection~\cite{leppakangas_multiplying_2018}. With increasing $n$, the position of this maximum shifts towards lower fluxes in the SQUID loop, thus higher Josephson energies. This behaviour is well explained by theoretical calculations~\cite{leppakangas_multiplying_2018}.

Ideally, at this point the conversion probability should be 1 and the reflection probability 0. For the $n=1$ case, panel (a), this is almost the case: At $\Phi=0.47\Phi_0$, the conversion probability reaches 0.90 while the reflection probability drops below $10^{-2}$. For $n=2$ at $\Phi=0.39\Phi_0$, the conversion probability reaches 0.73 while the reflection probability is still low at $0.018$. For $n=3$ at $\Phi=0.28\Phi_0$, the conversion probability reduces to 0.69 while the reflection probability increases to $0.20$, due to an increase and a shift of the minimum of reflection. This increase in reflection is likely due to the competition between the intended conversion process and spurious amplification processes as discussed above.

\begin{figure}
  \includegraphics[width=\columnwidth]{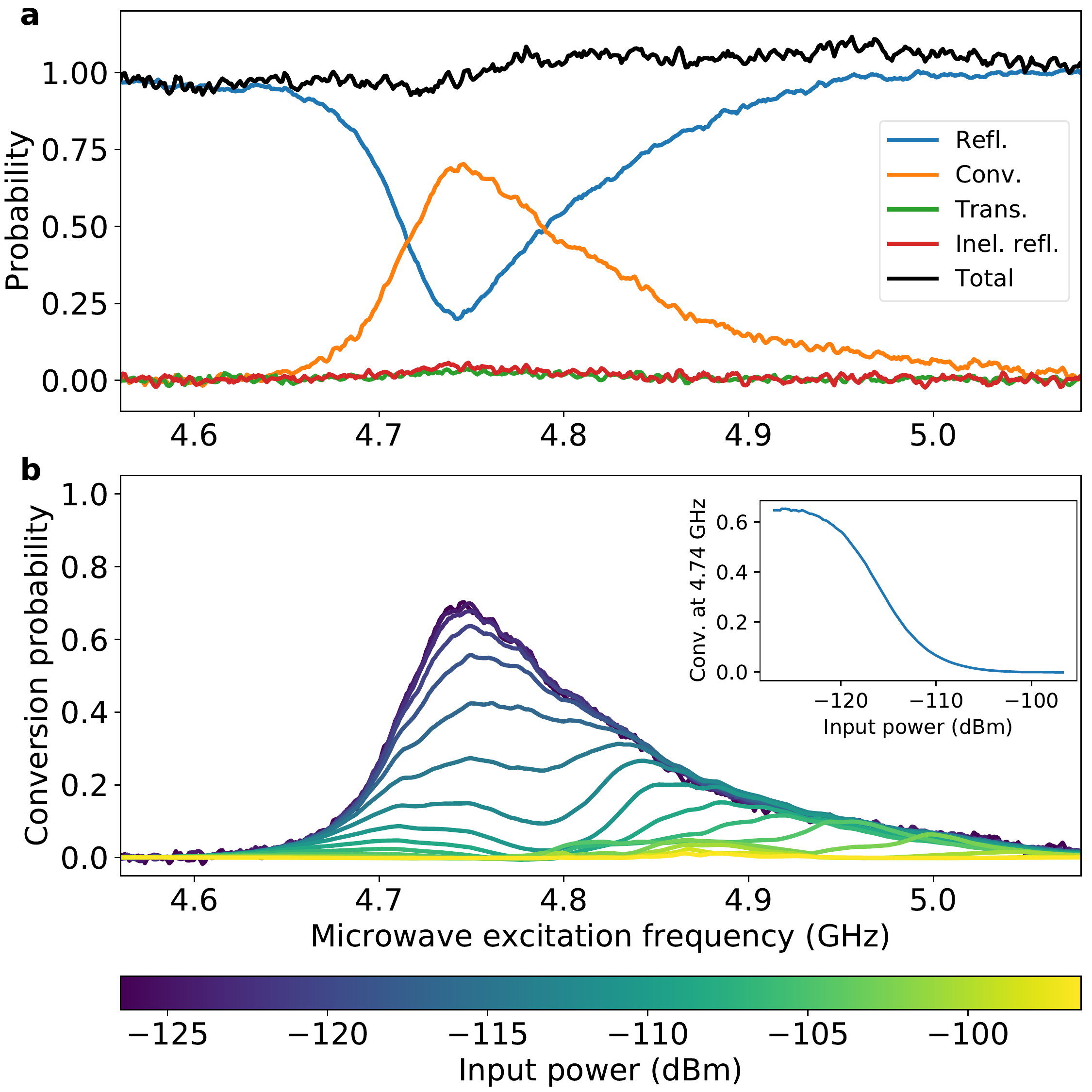}
  \caption{Bandwidth and saturation power.
  \textbf{a}, Reflection, conversion, transmission and inelastic reflection probabilities for the $n=3$ process, taken at the flux and voltage values maximising the conversion probability indicated by a dashed vertical line in Fig.~\ref{fig:impedancematching}c. The input power is $\SI{-127}{\deci\belm}$.
  \textbf{b}, Conversion probability for the $n=3$ conversion for various input powers. The inset shows the conversion probability at $\SI{4.74}{\giga\hertz}$ (maximum of conversion at low input power) as a function of input power.
    \label{fig:satpower}}
\end{figure}

By changing the input frequency $\nu$ as well as the input power, we can obtain the bandwidth and saturation power of the photo-multiplier. In Fig.~\ref{fig:satpower}, we show those results for $n=3$. The same curves for $n=1$ and $n=2$ can be found in App.~\ref{app:sat1and2}. Fig.~\ref{fig:satpower}a shows the reflection and conversion probabilities at low input power, below one photon on average in the input resonator. The maximal conversion efficiency is 0.69 and the full width at half maximum is $\SI{116}{\mega\hertz}$. This width is set by the bandwidth of the microwave resonators in the setup~\cite{leppakangas_multiplying_2018}.

In Fig.~\ref{fig:satpower}b the input power is gradually increased towards $\SI{-100}{dBm}$. The $\SI{1}{\deci\bel}$ compression point is $\SI{-119}{\deci\belm}$, corresponding to an approximate photon input rate of $\SI{400}{MHz}$. As the power increases, the shape of the curve changes drastically: it first splits in two maxima, then it acquires a third maximum. This behaviour arises from the non-linearity of the Josephson Hamiltonian which, in addition to the desired conversion term has higher order terms which are resonant at the same voltage, but modify the conversion rate as a function of power~\cite{leppakangas_multiplying_2018}. 

The main source of noise added by our device is the spontaneous emission of photons. When operated as a single photon detector, these emitted photons may be interpreted as false positive events, even though they have weaker bunching than the desired converted photons. This spontaneous photon emission rate is plotted in grey in Fig.~\ref{fig:impedancematching}. At the operating point of the $n=3$ conversion, it is $\approx\SI{400}{\mega\hertz}$.  It is mainly due to the large Josephson energy needed to efficiently perform the conversion for the output resonator with characteristic impedance $\approx\SI{400}{\Omega}$. As seen in Fig.~\ref{fig:spontaneous}c, at flux $0.28\Phi_0$ (operation point for the data of Fig.~\ref{fig:satpower}), the spontaneous emission background is dominated by the emission due to higher order processes involving several Cooper pairs and photons. A characteristic impedance closer to $h/(4\pi e^2)\sim\SI{2}{\kilo\ohm}$ or a lower bandwidth would require lower Josephson energy for optimal conversion~\cite{leppakangas_multiplying_2018}. A lower Josephson energy would in turn, as Fig.~\ref{fig:impedancematching}c shows, lead to significantly reduced spontaneous emission rate.

In summary, we have experimentally demonstrated multiplication of the photon number of an incoming microwave signal by an integer factor of 3, with efficiency reaching 0.69. The key characteristics of our device are well understood theoretically~\cite{leppakangas_multiplying_2018}. Given its compression point, our results could pave the way towards cascading two such devices to achieve a photon number gain of $9$, or even more. This gain would be sufficient to discriminate photon-numbers at the input by reading out the photo-multiplier with a quantum-limited amplifier \cite{leppakangas_multiplying_2018}, and thus detect and count single photons. Note that in such a scheme the observed spontaneous emission rate does not directly cause an equivalent dark count rate, because detected incoming photons create a bunch of $n$ photons while we expect the photons created spontaneously to come in smaller bunches depending on the stage at which they are created, so that the can be discriminated from incoming photons \cite{leppakangas_multiplying_2018}. Moreover the observed spontaneous emission rate can be significantly reduced by increasing the characteristic impedance of the output mode. It should then be possible to use cascaded multipliers as single photon counters without dead time and able to resolve photon numbers. Such a device would then implement a linear photon number amplifier with minimal added photon noise at the expense of losing, or at least deamplifying phase information.

  \textbf{Author contributions:} RA designed and fabricated the device based on an idea by MH and performed first measurements. JG performed the final measurements, analysed the data together with RA and NB, and performed simulations. All authors contributed to the experimental setup and calibration methods. FB together with RA, MH, UM and JG designed the software for performing the experiment. JG wrote the paper together with RA, MH and NB and input from all authors.

  \textbf{Acknowledgements:} This work was supported by the Natural Sciences and Engineering Research Council of Canada, the Canada First Research Excellence Fund, the European Union (ERC starting grant 278203 WiQOJo) and the french Agence Nationale de la Recherche (grant JosePhSCharLi).

\bibliographystyle{apsrev4-2}
\bibliography{photomultiplication-stage.bib}
\newpage
\appendix

\section{\label{app:fullsetup}Detailed experimental setup}

\begin{figure}
  \includegraphics[width=\columnwidth]{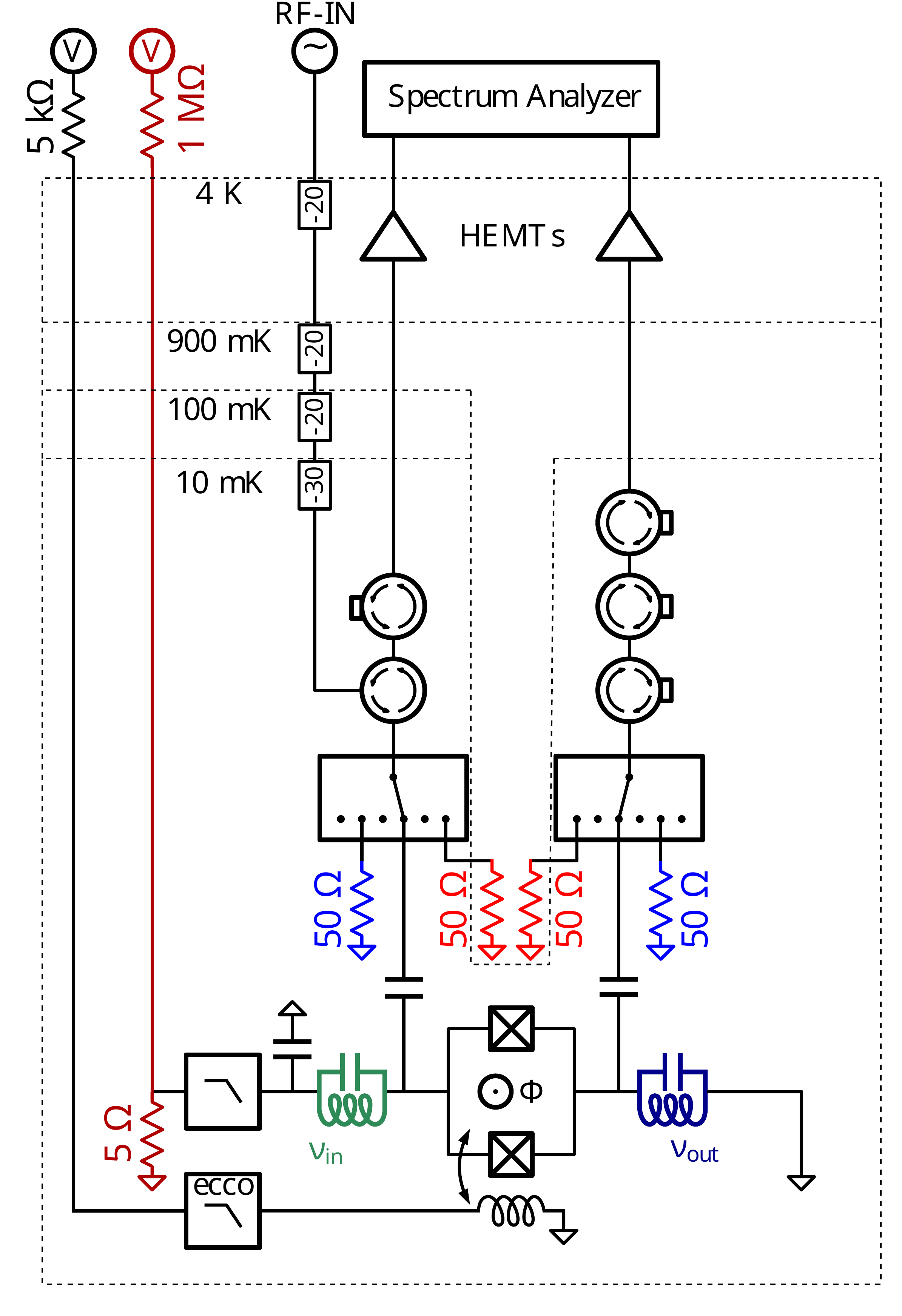}
  \caption{Schematics of the experimental setup in the dilution refrigerator.
    \label{fig:fullsetup}}
\end{figure}

The input and output port of our device are connected to six-port cryogenic switches allowing for in-situ calibration of the amplification chain (detailed in App.~\ref{app:calibration}). The common ports of the switches are connected to 0.3 to \SI{14}{\giga\hertz} cryogenic HEMT amplifiers with noise temperature $\Tn\approx\SI{3.5}{\kelvin}$ through a 4 to \SI{8}{\giga\hertz} double-junction circulator on the input side and a 5 to \SI{12}{\giga\hertz} triple-junction circulator on the output side. The input-side circulator is also connected to a microwave source at room temperature in order to send microwave signals into the device via line with $\SI{90}{\deci\bel}$ nominal attenuation. At room temperature, the amplified signals are down-converted into the $0-1$~\si{\giga\hertz} band via a custom double-heterodyne receiver, so that both input and output signals can be acquired at the same time by a $2\times \SI{2}{\giga\sample\per\second}$ ADC board.

A DC bias voltage can be applied to the sample via a voltage divider formed by a $\SI{1}{\mega\ohm}$ resistor at room temperature and a $\SI{5}{\ohm}$ resistor at base temperature. This voltage is then low-pass filtered with a custom filter combining discrete components and a silver-epoxy filter made with superconducting wire for high-frequency filtering (above $\SI{10}{\mega\hertz}$) and thermalisation. This filter is similar to the one used in~\cite{albert_multiplication_2019}. It has a cutoff frequency of \SI{700}{Hz} and a flat output impedance of \SI{5}{\ohm} up to several hundred $\SI{}{\mega\hertz}$. Combined with an on-chip \SI{100}{\pico\farad} capacitor which shorts the signal to the ground at the working frequency, the voltage bias circuit can, therefore, be modelled as as simple $RC$ circuit. 

The flux line for the SQUID is biased by a room-temperature voltage source in series with a $\SI{5}{\kilo\ohm}$ resistor. It is filtered at base temperature with a custom Eccosorb filter~\cite{paquette22} with a cut-off frequency of the order of $\SI{200}{\mega\hertz}$.

When measuring our device, we alternate between an ``on''-state PSD at the device parameters we want to measure and an ``off''-state PSD without voltage bias or input microwave tone. Calculating the difference of these two measurements allows as two accurately subtract the noise of our measurement setup. The PSDs are then calibrated as described in the next section.

\section{\label{app:calibration}Calibration of the amplification chain}
To obtain the actual power at the output of our device, we have to calibrate our measurement chain. To do so, for both measurement channels, one cold ($T_\mathrm{C}\approx\SI{10}{\milli\kelvin}$, blue in Fig~\ref{fig:fullsetup}) and one hot ($T_\mathrm{H}\approx\SI{900}{\milli\kelvin}$, red) $\SI{50}{\Ohm}$ resistor are successively connected to the switches. Their thermal noise is then acquired in the $4-\SI{8}{\giga\hertz}$ band. From these to measurements we calculate the gain of the measurement chain from the switches to the digitiser (Y-factor calibration). To ensure correct thermalisation of these resistances, they are thermally isolated from the switches via superconducting NbTi coax lines and thermally anchored to, respectively, the mixing chamber stage and the still stage of our dilution refrigerator.

To calibrate the input line, we perform a reflection measurement with the input switch either on an open port, or connected to the cold $\SI{50}{\Ohm}$ resistor. This allows us to correct for leakage of the circulators, and gives us the gain of the input line down to the input switch.

We apply a last layer to this calibration by measuring the reflection and the transmission of our device when the bias voltage is far from any working point and the SQUID maximally frustrated. By doing so, we can consider the device as a non-dissipative linear component which either reflects or transmits the input signal (at least out of the input resonator bandwidth). This is used to eliminate the contribution of the last cable from the switch to the sample, which we model as a constant attenuation over our frequency window to remove the impact of the input resonator. This correction amounts to $\SI{0.267}{dB}$, in agreement with the expected properties of the cable and other measurements done on different samples. This ultimately allows compute at each frequency the actual input and output photon rates at the device, which we use to calculate the different probabilities discussed in this work.

\section{\label{app:vnoise}Bias voltage noise}

\begin{figure}
  \includegraphics[width=\columnwidth]{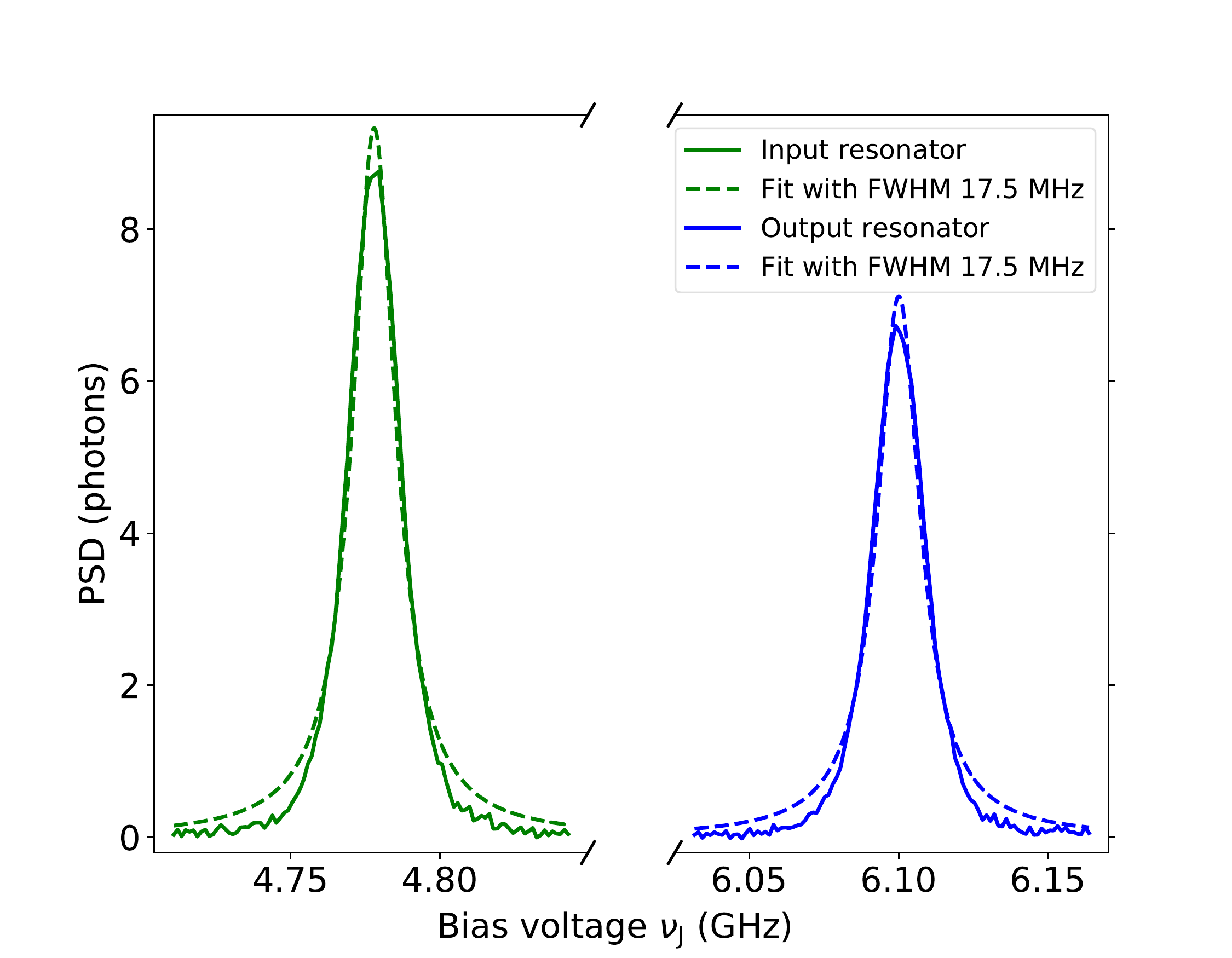}
  \caption{Voltage noise. PSD of the input resonator at $\SI{4.78}{\giga\hertz}$ (green) and  of the output resonator (blue) at $\SI{6.1}{\giga\hertz}$, both taken at $\Phi=\Phi_0/2$.
    \label{fig:vnoise}}
\end{figure}

Fig.~\ref{fig:vnoise} shows the PSD of the input and output resonators at maximal SQUID frustration, measured at the resonance frequency of the resonators for a voltage around the process where one Cooper pair gives one photon. In both cases, the width of the measured Lorentzian is $\SI{17.5}{\mega\hertz}$.

According to the $P\left(E\right)$ theory~\cite{ingold92, albert_multiplication_2019}, this full width at half maximum $\gamma$ is given by the thermal noise of the low-frequency electromagnetic environment, in this case the bias resistor $R_b=\SI{5}{\ohm}$ at electronic temperature $T_e$:
$$\gamma = \frac{2 R_b}{\hbar R_Q}k_BT_e.$$

This allows extracting the effective electronic temperature of the bias resistor. We find $T_e=\SI{86}{\milli\kelvin}$. However, this value has to be taken with caution, and is likely significantly overestimated because for $P(E)$ to be valid, the electromagnetic environment must stay in thermal equilibrium, which is not a good approximation here: Fig.~\ref{fig:vnoise} shows that even at maximal frustration of the SQUID, the PSD on resonance is larger than 1 photon, a strong deviation from thermal equilibrium. Under this conditions, we expect the Lorentzians to be compressed, leading to an overestimation of the temperature of the bias resistor. On devices with lower Josephson energy we indeed observe linewidths corresponding to $T_e\approx\SI{20}{\milli\kelvin}$ to $\SI{30}{\milli\kelvin}$.

\section{\label{app:impmatching}Response as function of bias voltage}

\begin{figure}
  \includegraphics[width=\columnwidth]{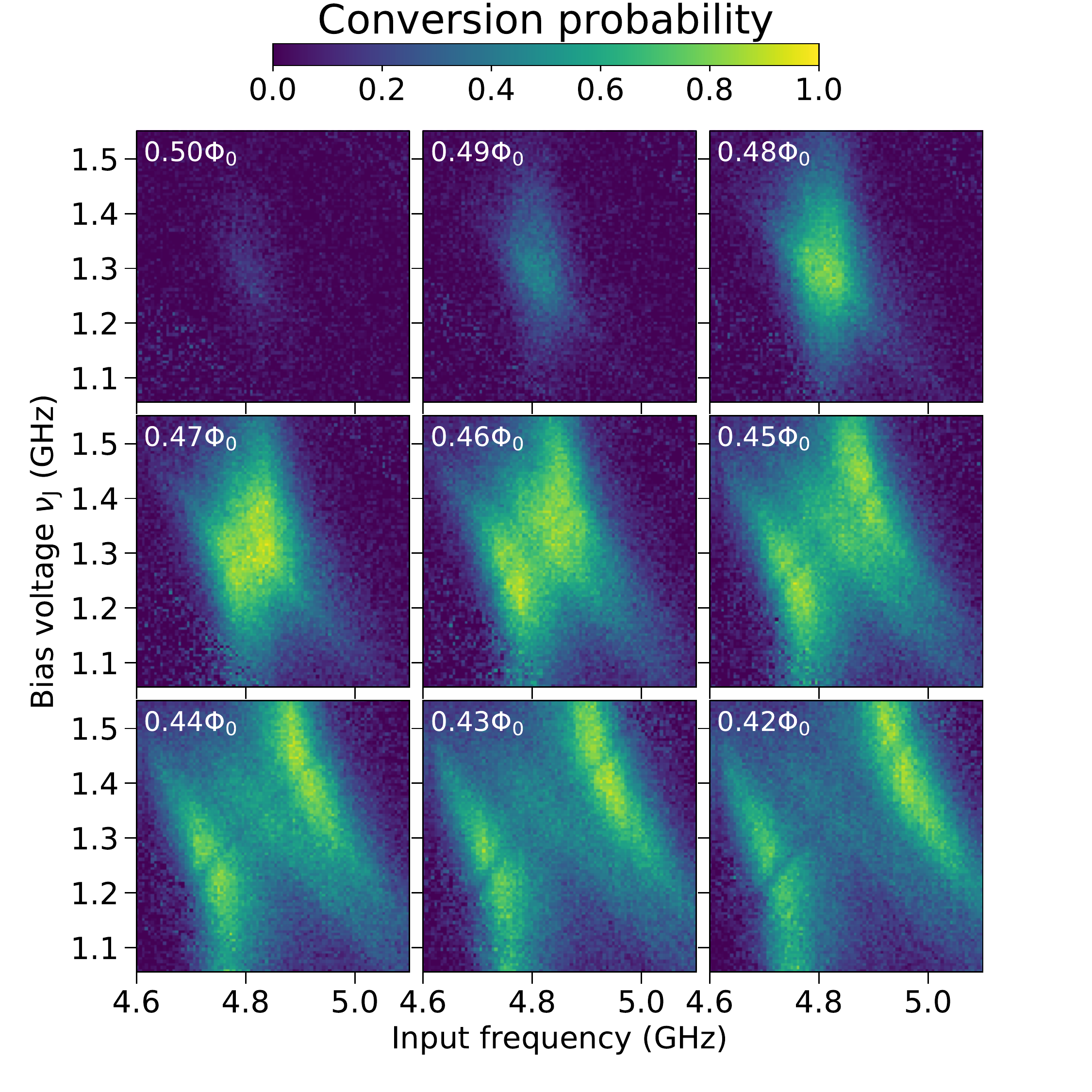}
  \caption{
  Measured conversion probability for the $n=1$ conversion as a function of input frequency and bias voltage for several SQUID flux values. $0.47\Phi_0$ corresponds to the optimal operation point represented by a dashed vertical line in Fig.~\ref{fig:impedancematching}a.
    \label{fig:appimpmatch}}
   \includegraphics[width=\columnwidth]{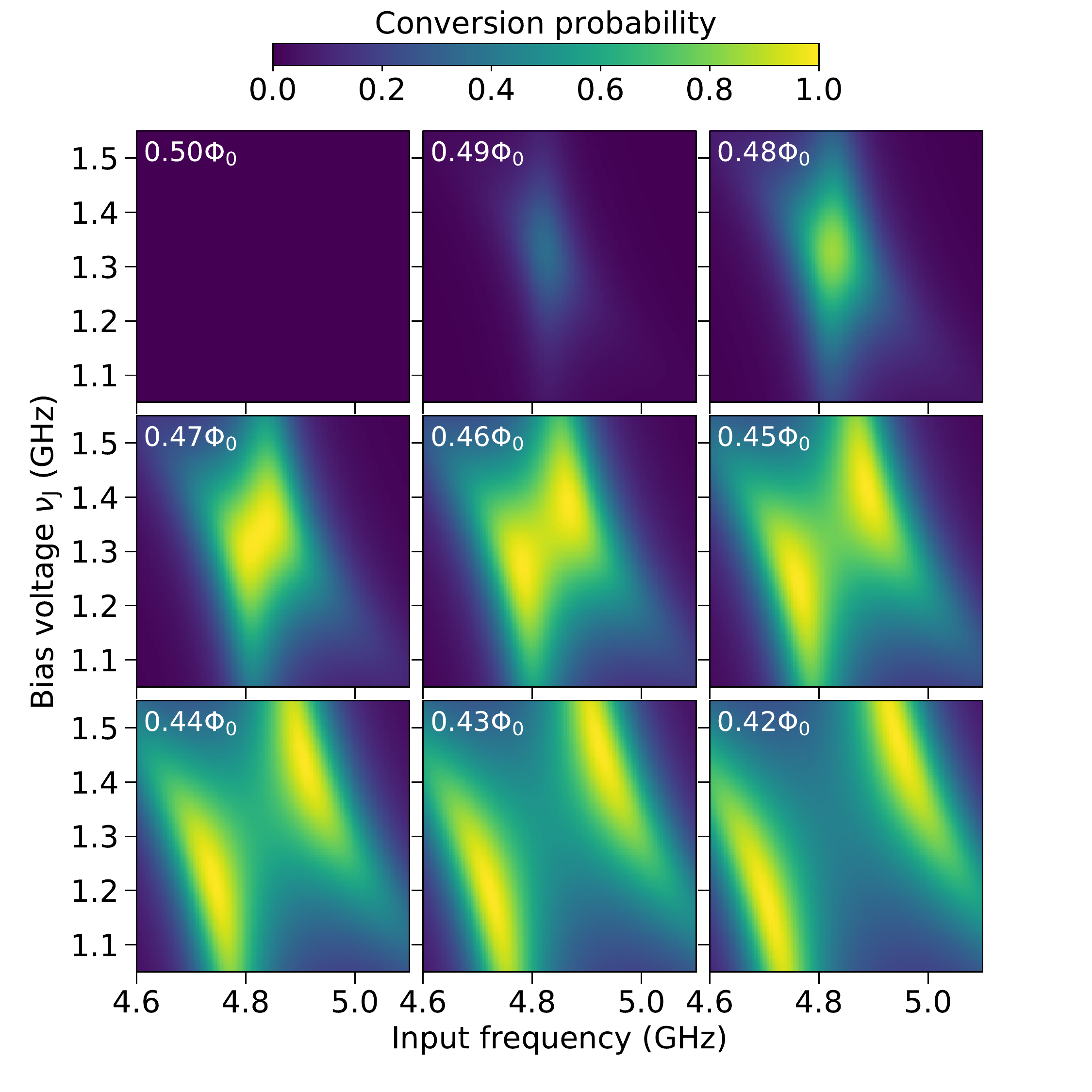}
  \caption{
  Fit of the conversion probabilities for $n=1$ from Fig.~\ref{fig:appimpmatch} to theory \cite{leppakangas_multiplying_2018}. The fit is performed at the optimal SQUID frustration $0.47\Phi_0$. All other SQUID fluxes use the same fitting parameters.
    \label{fig:appimpmatchsimu}}
\end{figure}

In the main text, we show the response of our device only at the optimal bias point as a function of the flux in the SQUID loop in Fig.~\ref{fig:impedancematching}, or the frequency in Fig.~\ref{fig:satpower}. Fig.~\ref{fig:appimpmatch} shows the conversion probability for the $n=1$ conversion as a function of the input frequency and the bias voltage for several flux values. For values above $0.47\Phi_0$ (Josephson energies below the optimal value), there is one maximum, the amplitude of which increases when the flux approaches $0.47\Phi_0$. Below this value, there are two maxima because the input and output modes enter the strong coupling regime. As SQUID bias is further reduced, the splitting in the anticrossing increases.  This behaviour is well described by input-output theory applied to the circuit~\citep{leppakangas_multiplying_2018}. The dark line with slope 1 (visible at fluxes below $0.46\Phi_0$) in the data of Fig.~\ref{fig:appimpmatch} corresponds to a spurious mode in the junction environment at $\SI{3.5}{\giga\hertz}$.

Fig.~\ref{fig:appimpmatchsimu} shows a fit of the conversion probability for the $n=1$ conversion. These curves depend on the Josephson energy and on the resonance frequency and width of the input and output modes~\cite{leppakangas_multiplying_2018}. Given the assumption that a flux of $0.47\Phi_0$ puts the Josephson energy at its ideal matching value, this removes the Josephson energy parameter, and we use that point to perform our fit. We calculate the remaining 2D conversion plots using the results from that fit, and they reproduce well our experimental results. The calculated values differ from the measured ones close to full frustration, likely because in the calculation we assume a perfectly symmetric SQUID, whereas the actual SQUID is not perfectly symmetric as can be seen in the experimental data for $0.5\Phi_0$. The extracted resonance frequencies match well the previously obtained frequencies. The obtained width of the input mode also matches the designed coupling rate, with $\gamma_\mathrm{in}^\mathrm{fit}\approx \SI{90}{MHz}$. However, the result for the width of the output mode $\gamma_\mathrm{out}^\mathrm{fit}\approx \SI{220}{MHz}$ differs significantly from the design value, with. This discrepancy might indicate that the theoretical model is missing part of the behaviour of our system. This behaviour was also observed in earlier measurement on similar devices~\cite{albert_multiplication_2019} and could also be related to the spurious processes observed in Fig.~\ref{fig:spontaneous}, or to the change of optimal input frequency observed in Fig.~\ref{fig:impedancematching}.

%\section{\label{app:nmax}Maximal number of photons at the input}
%
%\begin{figure}
%  \includegraphics[width=\columnwidth]{FigureSM_1to3.pdf}
%  \caption{
%  Relative error made on measurement.
%    \label{fig:appnmax}}
%\end{figure}
%
%Fig.~\ref{fig:appnmax} shows the relative error made on estimating the number of photons at the input of the photo-multiplier device by measuring this number at the output in the case of the one to three conversion. The blue curve is calculated for each input power (expressed as a number of input photons $n_\mathrm{max}$ in the input resonator) of Fig.~\ref{fig:satpower}b) by comparing the measured conversion probability at $\SI{4.74}{\giga\hertz}$ (maximum of conversion at low input power) to the maximal value of 0.75. The orange curve is given by $1/n_\mathrm{max}$ and corresponds to \note{?????}

\section{\label{app:sat1and2}Saturation power for $n=1$ and $n=2$}

\begin{figure}
  \includegraphics[width=\columnwidth]{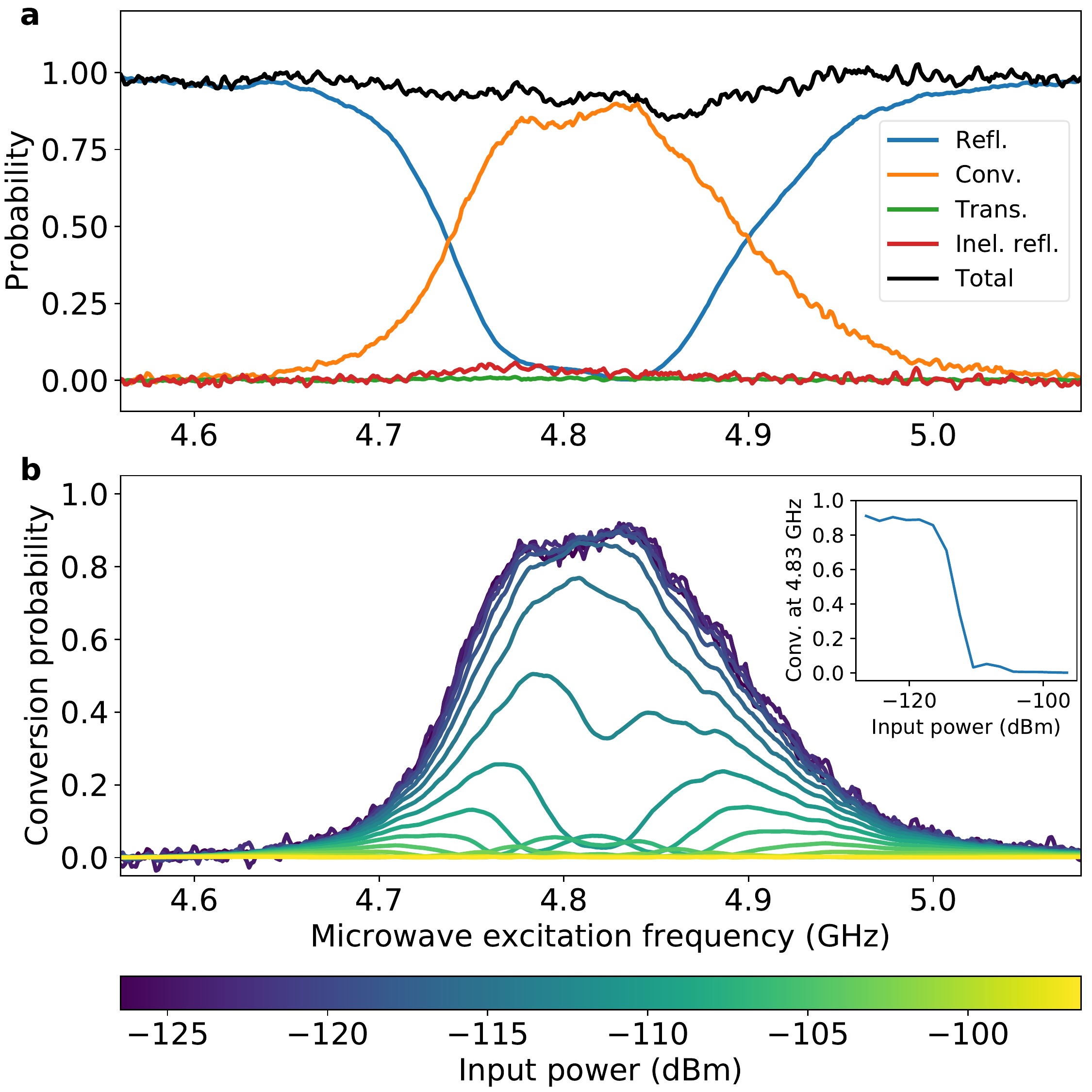}
  \caption{Bandwidth and saturation power for $n=1$ conversion. \textbf{a}, Reflection, conversion, transmission and inelastic reflection probabilities taken at the flux value maximising the conversion probability, indicated by a dashed vertical line in Fig.~\ref{fig:impedancematching}a. The input power is $\SI{-127}{\deci\belm}$. \textbf{b}, Conversion probability for various input powers. The inset shows the conversion probability at $\SI{4.84}{\giga\hertz}$ (maximum of conversion at low input power) as a function of input power.
    \label{fig:satpower1to1}}
\end{figure}

\begin{figure}
  \includegraphics[width=\columnwidth]{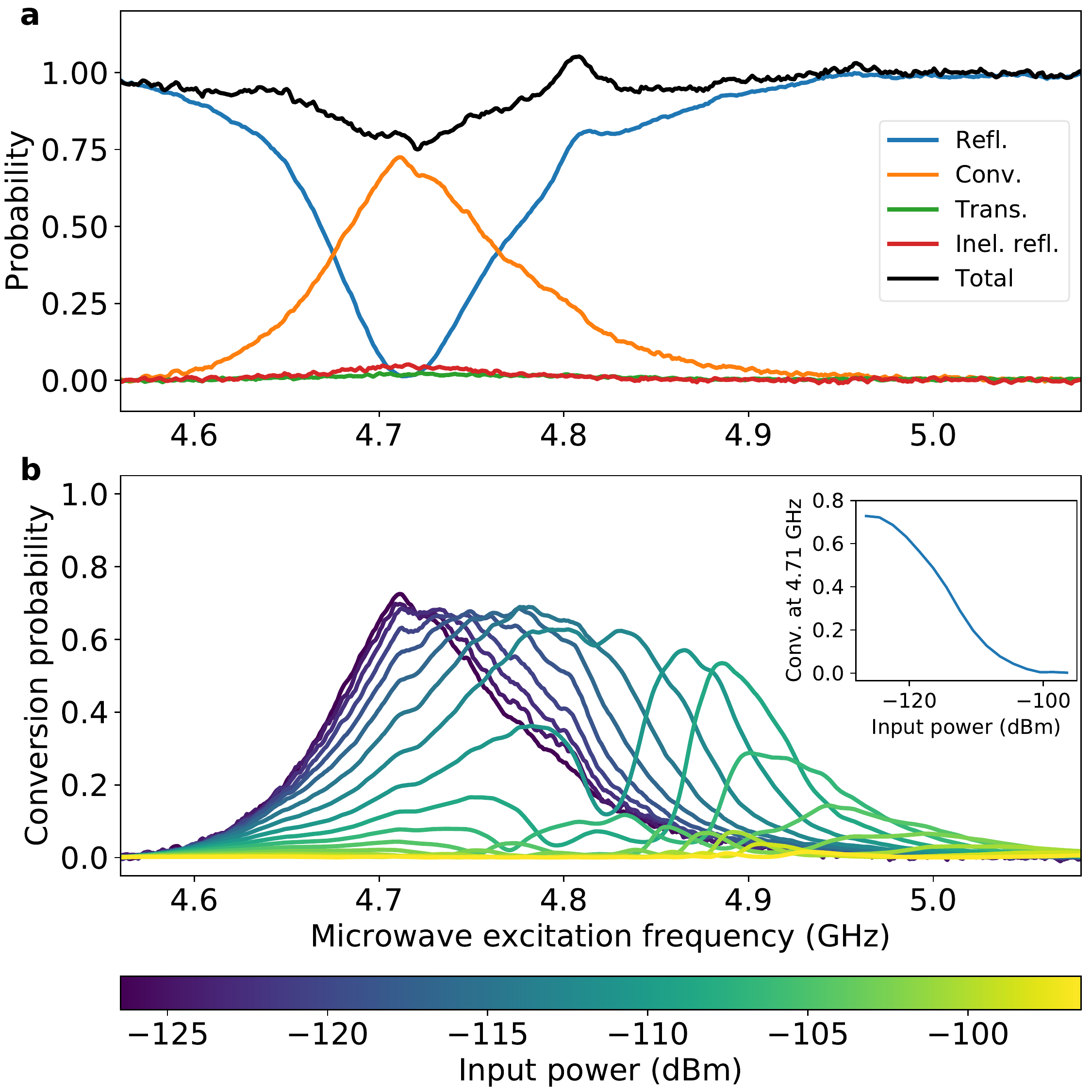}
  \caption{Bandwidth and saturation power for the $n=2$ conversion. \textbf{a}, Reflection, conversion, transmission and inelastic reflection probabilities taken at the flux value maximising the conversion probability, indicated by a dashed vertical line in Fig.~\ref{fig:impedancematching}b. The input power is $\SI{-127}{\deci\belm}$. \textbf{b}, Conversion probability for various input powers. Inset shows the conversion probability at $\SI{4.71}{\giga\hertz}$ (maximum of conversion at low input power) as a function of input power.
    \label{fig:satpower1to2}}
\end{figure}

Fig.~\ref{fig:satpower1to1} and~\ref{fig:satpower1to2} show the bandwidth and saturation power of the photomultiplier for the $n=1$ and $n=2$ conversion.

The (a) panels of both figures are taken at low input power, close to one photon on average in the input resonator. They represent the measured reflection, conversion, converted reflection and transmission probabilities. For the $n=1$ conversion, the maximal conversion probability is 0.90, obtained at $\SI{4.84}{\giga\hertz}$. At this frequency, the reflection probability drops below $10^{-2}$, which corresponds to the directivity of our circulators. The bandwidth of this process is $\SI{163}{MHz}$. The total measured signal (black line) stays close to 1 in the whole measurement bandwidth showing that no other process is taking place at the same time. For the $n=2$ conversion, the maximal conversion only reaches 0.73 at $\SI{4.71}{\giga\hertz}$ with a reflection dropping to 0.018. The bandwidth of this process is $\SI{106}{MHz}$. The total measured signal has a small dip at this frequency. This reflects the fact that either the junction is emitting outside of the measurement bandwidth or that power is dissipated in the setup.

In panels (b), the input power is increased to a few hundred photons on average in the input resonator and the conversion probability is plotted for each input power. The same behaviour as for the one to three conversion (Fig.~\ref{fig:satpower}b) is observed for the one to one conversion (Fig.~\ref{fig:satpower1to1}b): Several maxima appear as the power is increased. The $\SI{1}{\deci\bel}$ compression point for the one to one conversion is $\SI{-114.5}{\deci\belm}$, corresponding to an input photon rate of approximately $\SI{1.1}{GHz}$. The one to two conversion (Fig.~\ref{fig:satpower1to2}) has a slightly different behaviour: the frequency at which the maximal conversion probability is observed is shifting to higher frequencies as the input power is increased. The $\SI{1}{\deci\bel}$ compression point for the $n=2$ conversion is $\SI{-118.5}{\deci\belm}$, corresponding to an input photon rate of approximately $\SI{440}{MHz}$. Those values are larger than for the $n=3$ conversion presented in the main text. This is not surprising as the energy in the output mode increases with $n$ for a given input power.

\end{document}